# Effective Influence Spreading in Temporal Networks With Sequential Seeding

RADOSŁAW MICHALSKI[1,2], (Member, IEEE), JAROSŁAW JANKOWSKI[2], AND PIOTR BRÓDKA[1,2]
[1]Department of Computational Intelligence, Wrocław University of Science and Technology, 50-370 Wrocław, Poland
[2]Department of Computer Science and Information Technology, West Pomeranian University of Technology, 71-210 Szczecin, Poland

Corresponding author: Radosław Michalski (radoslaw.michalski@pwr.edu.pl)

This work was supported by the National Science Center, Poland, under Grant 2016/21/B/HS4/01562.

**ABSTRACT** The spread of influence in networks is a topic of great importance in many application areas. For instance, one would like to maximise the coverage, limiting the budget for marketing campaign initialisation and use the potential of social influence. To tackle this and similar challenges, more than a decade ago, researchers started to investigate the influence maximisation problem. The challenge is to find the best set of initially activated seed nodes in order to maximise the influence spread in networks. In typical approach we will activate all seeds in single stage, at the beginning of the process, while in this work we introduce and evaluate a new approach for seeds activation in temporal networks based on sequential seeding. Instead of activating all nodes at the same time, this method distributes the activations of seeds, leading to higher ranges of influence spread. The results of experiments performed using real and randomised networks demonstrate that the proposed method outperforms single stage seeding in 71% of cases by nearly 6% on average. Knowing that temporal networks are an adequate choice for modelling dynamic processes, the results of this work can be interpreted as encouraging to apply temporal sequential seeding for real world cases, especially knowing that more sophisticated seed selection strategies can be implemented by using the seed activation strategy introduced in this work.

## I. INTRODUCTION

Information diffusion and the spread of influence within social networks attracts the attention of researchers from various fields for many years. Initially, the studies have been focusing on identifying the factors or personal attributes that facilitated these processes, and were based on small scale experiments. However, with the increase of computational power and data availability, new possibilities emerged that allow to investigate these phenomena in large scale. This resulted with new research areas and sub-disciplines, such as network science [1], computational network science [2], and computational social science [3].

While influence spreading processes within complex networks are observed in various areas, most studies focus on their modelling [4], coverage prediction [5], analysis of interacting processes [6], increasing of their dynamics or

The associate editor coordinating the review of this manuscript and approving it for publication was Chun-Hao Chen.

suppressing negative activity [7]. In the area of the spread of influence, early approaches focused on the selection of seeds for a given influence model. These seeds, if selected optimally, when influenced (activated) at the beginning of the process, allow to receive the highest outcome regarding the number of activated nodes at the end. The challenge on how to select seeds was posed by Kempe *et al.* [8]. There was a given budget $k$ for activating the nodes, the cost of activation has been the same for all nodes and the network has been static. This research direction resulted in several approaches - starring with the most effective greedy approach [8] and its extensions with adjustable computational performance [9]. Other seed selection methods were also explored, including heuristics based on centrality measures [10], community seeding [11], [12], k-shell decomposition [13] and other solutions [14]–[18].

Initial studies in the area of influence maximisation focused mainly on static networks. However, over time they progressed towards more realistic scenarios, like dynamic networks. The first influence maximisation method for dynamic networks was based on adding time factor to classical influence maximisation problem [19]. In this work, the mechanism for seed identification based on backward and forward influence algorithms was proposed. It was further explored and compared against a static approach in [20], where authors found out that by using a dynamic network and by considering the changes in the network in the process of seed selection, they were able to find a better seed set such that the final set of influenced nodes was larger. By using this seed set, the spreading process activated two times more nodes than the seed set built by basing on a static network. Other alternative approaches proposed finding influential seed successors in social networks [21] and compensatory seeding, taking into account the availability of nodes within the network [22].

Most of existing solutions are based on the selection of seeds in a single step and the initialisation of the process as a single stage. After the first step of initialisation we can only observe the process without the possibility to influence it anyhow. However, from real world scenarios we know that usually it is better to split our decision into a series of smaller ones and stretch them in time so that each subsequent decision is supported by additional information. The examples from variety of fields, like theory of decision making [23], financial markets [24], epidemiology [25] or market strategies [26] show that changing a single decision into a sequence of subdecisions reduces risk, allows to gather knowledge about the consequences of the earlier ones, and enable the use of additional information that is revealed as time passes. As a result, this knowledge can be exploited in later stages for better decision-making. The same situation relates to seeds activation. Activating all of them at the beginning may lead to situation when nodes activated as seeds could be activated in a natural process by other nodes. Thus, sometimes it might be better to wait and see which nodes will be activated by initial seeding with only part of our budget $k$, to see if we cannot spend the rest of it better.

Based on this idea, new approaches were proposed in a form of adaptive seeding and heuristics like seeding scheduling [27], active marketing [28] and usage of gathered knowledge for Linear Threshold Model [29]. The advantage of distributing seeds activation over the time in most general form has been explored by the sequential seeding [30]. This approach, described in more detail in next Section, proved its superiority over the activation of seeds in single stage in static networks. Instead of using all seeds at the beginning, only a part of them is used and the rest is saved for later to revive the process after it stops or slows down. The study was based on the highest possible decomposition of the problem for independent cascade model [31], with only one seed used per stage. It was demonstrated that it increases the coverage for any used seed selection approach, if the seeds are activated sequentially instead of in a single stage. The results proved that this approach yields at least the same results as a ''traditional'' single stage seeding for the same conditions [32]. This is especially the case for models that base on independent activations, since for threshold-like models, e.g. linear threshold having a committed neighbourhood is required to activate a node and different strategies should be applied there, as shown in [33], [34].

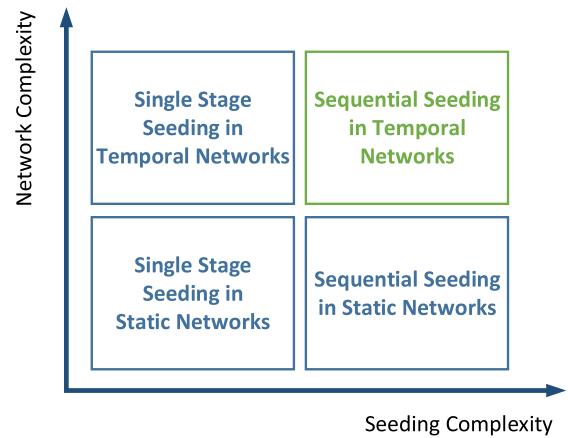

**FIGURE 1.** The landscape of research on seeding approaches. The boxes in blue are the areas that are actively studied, i.e., single stage seeding in static networks ([8]–[18]); single stage seeding in temporal networks ([19]–[22]), and sequential seeding in static networks ([27]–[30], [32], [35]–[39]) and the one in green is explored in this work, namely sequential seeding in temporal networks.

The evolution of seed maximisation approaches described in this section has been shown in Fig. 1. The rectangles in blue are showing which areas are actively studied so far and the one in green moves us to the challenge tackled in this work, namely the exploration of the process of seed distribution in a sequential way in temporal networks. While earlier studies proved the performance of sequential seeding, the structures of used networks were static. Changes of networks in terms of nodes and edges create another analytical dimension and new challenges. This makes the problem closer to reality with changing structures over the time. This work introduces the sequential seeding method for temporal networks and the verification of its performance is one of the main goals. The experimental setup is based on a set of real networks with temporal characteristics and independent cascades spreading model adjusted for the temporal setting. The results demonstrate improved performance when compared to single stage seeding in most simulation cases with results dependent mainly on activation probabilities of the influence model.

This work is organised as follows. Section II provides an introduction to sequential seeding approach with illustrative examples regarding seeding in temporal networks. Section III introduces the methods used in this work alongside with experimental space parameters. Section IV shows results from simulations performed within datasets based on real networks and contains the discussion on them. Section V summarises the findings and proposes future work directions.

## II. SEQUENTIAL SEEDING FOR TEMPORAL NETWORKS

The sequential seeding approach takes into account several scenarios. It bases on an unconditional seeding when seeds are allocated to every stage of the process and are used in sequential steps without any assumptions on the state of the influence process. This leads to higher gains when the process has time to develop and initial superiority of single stage approach is weakened over iterations. The generic approach used by this method is shown in Fig. 2. As mentioned in Section I, sequential seeding is mostly suited for models based in independent activations rather than for threshold-like models, so most of the development in the area of sequential seeding applies to the family of models based on independent cascades.

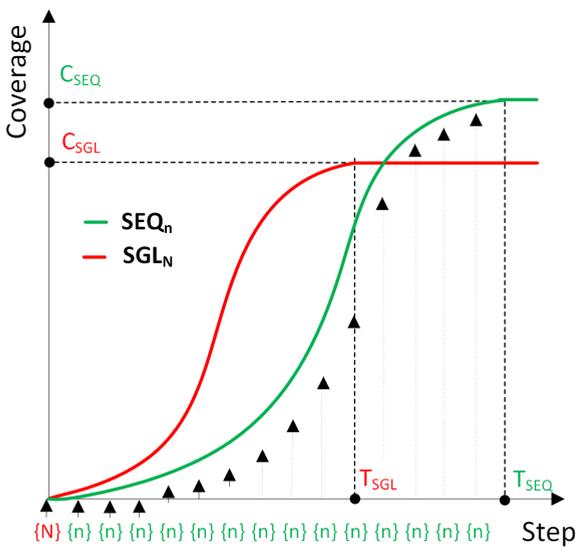

**FIGURE 2.** An illustrative example intended to demonstrate how Sequential Seeding (SEQ) can gain additional advantage over Single Stage seeding (SGL). Here, we compare sequential seeding activating $n$ seeds per stage with Single Stage seeding based on $N$ seeds used at the beginning to initiate the spreading process. Network coverage $C_{SEQ}$ represented by a fraction of activated nodes for Sequential Seeding is expected to be higher that coverage of Single Stage Seeding $C_{SGL}$ while it takes more time and $T_{SEQ} > T_{SGL}$.

More advanced solutions involving sequential seeding can be based on a balanced approach and a trade-off between the process duration and the final coverage. For example, if more than one seed per stage is activated, the coverage can be smaller but the duration of the process is reduced, what was showed for multiple seeds used in each step [30]. It was also proved that the sequential seeding provides at least the same coverage as a single stage seeding and is never worse for the same initial conditions [32]. Experimental setup was based on the coordinated execution approach with propagation probabilities assigned to network edges with possibility to test different methods within the same conditions, as described in Section III-D.

Sequential seeding was further extended towards the recalculation of rankings with effective degree used in each seeding stage [36]. In another study, the ability of buffering not used seeds was analysed with seeds collected in buffer during no seeding stages. Buffer was released as soon as process stops [35]. The effect of frequency of re-computations on final coverage was also analysed. Another studies in this area focused on the relationship between the network topology and the perfromance of sequential seeding [37]. Seed selection methods based on entropy centralities where used [39]. Another direction was the analysis of the effects of distribution of seeds over the time and usage of number of seeds following Gaussian, linear or geometric distributions [38].

Since in all earlier works the sequential seeding approach was designed and explored only for static networks, in this work we propose sequential seeding for temporal networks. The fact that temporal networks are different from static networks, since they evolve over time, introduces another perspective for sequential seeding. As it is not guaranteed that all seed nodes will appear in the future, the seeding budged has to be spent even more efficiently. This is the reason why we decided to evaluate how a sequential seeding method, that proved its superiority over single stage seeding for static networks, will perform in a temporal setting. However, firstly we need to discuss what are the changes introduced by temporal approach.

The first major difference is that the network changes over time, i.e. nodes can appear and disappear, the same can apply to edges. As a result of that, the seed set is being build by observing the network to a given point in time that could be considered as a training period. Next, for sequential seeding, the activations of seeds are being distributed over time considered as an evaluation period. Contrary to that, in a single stage seeding we activate all the nodes at ones at the beginning of spreading. Naturally, the model of a temporal network implies how distribution of seeds activations will be actually performed.

One of the most popular temporal network models is the model based on network snapshots. In this case all the events that belong to a certain period constitute a network snapshot representing the activities in this period. These snapshots are then time-ordered and this allows us to perform sequential seeding in such a way that we distribute the activations over particular snapshots. It is worth underlining that such a model for temporal network allows to appear and disappear nodes and edges over time what reflects reality more than a model in which the set of nodes is known in advance. The detailed illustrative description of how single stage seeding and sequential seeding work in a temporal setting has been presented in Figure 3 and Figure 4, respectively. More details on the temporal network model are to be found in Section III-B.

## III. METHODS

In order to verify the performance of sequential seeding in a temporal scenario, a series of experiments has been conducted. In the subsections below we present the crucial elements of the experiment and Tab. 1 shows the configuration space of the simulated diffusion processes.

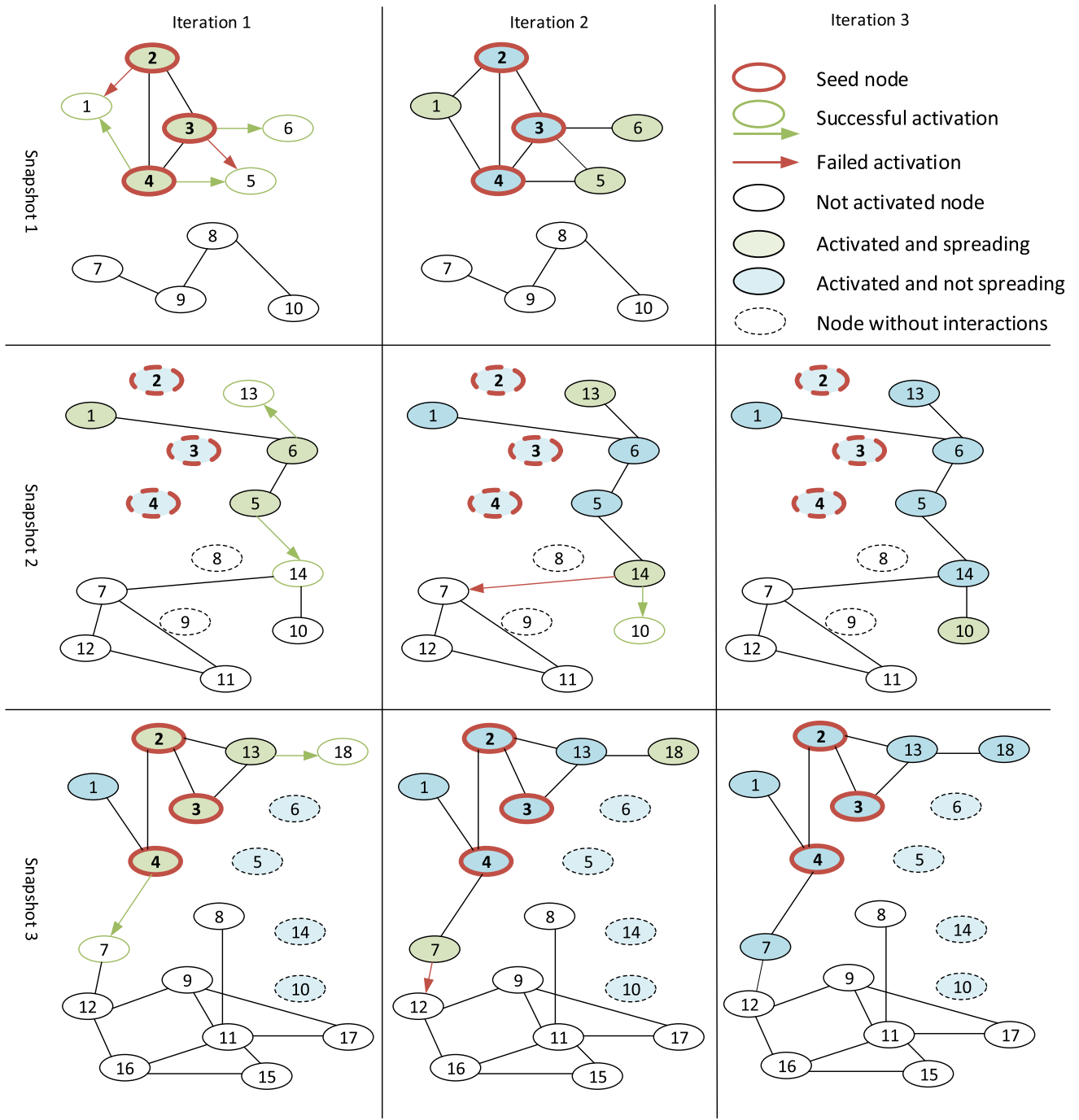

**FIGURE 3.** An example of "traditional" single stage seeding process in temporal network. In the rows we have three consecutive snapshots ($T_1$, $T_2$, $T_3$) of temporal networks. In the first snapshot ($T_1$) based on the current network information the degree centrality is calculated and top three nodes with the highest value of degree centrality are selected as seeds (nodes 2, 3 and 4). Seeds are activated and the spreading process begins. In each iteration depending on propagation probability, *p*, each activated node have one chance to activate each of its neighbours. Process ends when there are no nodes which can be activated or there are no nodes which can still activate its neighbours. After some time the network evolves/changes and the next snapshot ($T_2$) is created in which new nodes have appeared in the network (11, 12, 13, 14), what is more since nodes activities have been different during that time, also the connections between existing nodes have changed. This opens new possibilities for the spreading process since there are new neighbours which have not been tried before. Thus, the process once again progress until it cannot progress anymore. Similar situation is with the third snapshot ($T_3$). Finally, we end up with 11 (61%) activated nodes.

## A. DATASETS

The experiments have been conducted using two datasets: (i) manufacturing company email communication [40] consisting from emails sent between employees of a manufacturing company over a course of nine months and (ii) a Haggle dataset representing contacts between people measured by carried wireless devices [41]. Following [42], in order to make the results more independent from datasets, apart from performing experiments using original datasets, we did perform a rewiring procedure. It alters the aggregate network topology, yet preserves temporal structure locally on each link. The rewiring procedure has been performed on source datasets. We kept the events' times intact, but rewired the links by assigning different vertices to them. This is why the results are presented for both: non-rewired and rewired temporal networks. Manufacturing company dataset contains 167 nodes and 82,927 edges, Haggle dataset has 274 nodes and 28,244 edges. The choice for evaluated networks has been limited to mid-sized ones by the high number of experimental repetitions we decided to run following the results on the convergence of the independent cascade model presented in [43].

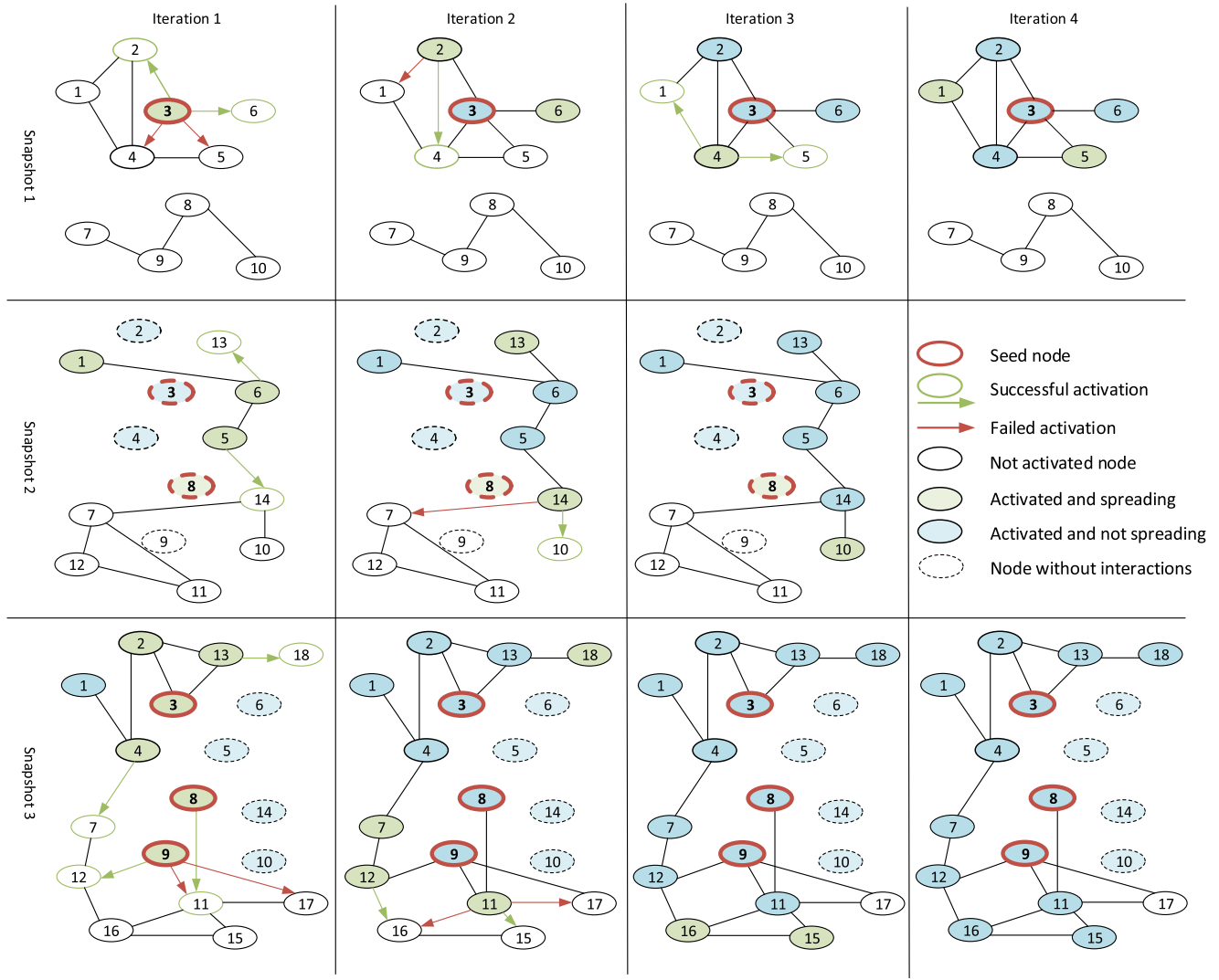

**FIGURE 4.** An example of sequential seeding in temporal network. In the rows we have three consecutive snapshots ($T_1$, $T_2$, $T_3$) of temporal network. In the first snapshot based on the current network information the degree centrality is calculated. What is different in compassion to single stage seeding presented on figure 3 we only use/add one seed node in each snapshot. Thus, in the firs snapshot ($T_1$) we select the one with the highest degree i.e. node 3, and start the spreading process in exactly the same way as in single stage seeding. Please note that using just one seed the spreading process activated the same number of nodes in the first snapshot ($T_1$) as single stage seeding (Fig. 3) using all three seeds, however it took more time, four iterations instead of three. Thanks to this approach we still have budget/resources for two more seeds, so in snapshot two ($T_2$) we are selecting as a seed the first not activated node on the ranking list (node 8). Unfortunately, since node 8 did not have any activities in this period it has no neighbours and cannot activate anyone, so the independent cascade model in snapshot two progress exactly the same as in snapshot two for single stage seeding. This shows the next weakens of sequential seeding which is that we are using the "historical" degree ranking from the first snapshot. This weakens can be easily addressed by recalculating rankings before each seeding as it was done for static networks [36]. However, since this approach additionally improves the sequential seeding approach we have decided not to do this in order to be able to compare the results of sequential seeding and single stage seeding. Despite the fact that node 8 could not activate anyone in the snapshot two it will be able to do so when it becomes active in the network like in snapshot three ($T_3$) in our example. In snapshot three node 8 together with newly activated third seed (node 9) are the foundation of activation cascade which activates most of the nodes not activated by the single stage seeding. Finally, we end up with 17 (94%) activated nodes.

It is worth underlining that even the networks that look relatively stable, can undergo changes that have significant impact on the network structure. For instance, for manufacturing company email dataset there has been one managerial position change that resulted with different information flow among multiple employees. In Section III-B we also describe what is the stability of nodes and edges across the snapshots for a given network configuration to demonstrate how varying temporal networks can be.

### B. TEMPORAL SOCIAL NETWORK

The temporal social network that is used for evaluation of sequential seeding is based on time windows (network snapshots). For all the evaluated datasets, the periods they cover have been split into *n* windows of equal size [44]. Then, all the events within a particular window became a source for building a temporal social network snapshot $T_n = (V_n, E_n)$, $n \in 1 \ldots N$ consisting from the set of nodes $V_n$ active in $n^{th}$ window and the set of undirected unique edges $E_n$ representing interactions between nodes in $n^{th}$ window. Each snapshot $T_n$ can be then considered as a static graph where all events between node *i* and *j* ($j, \in V_n$) are collapsed to the single edge $(i, j) \in E_n$. Since the snapshots are time-ordered, temporal aspects are preserved at a certain level of granularity. The reasons why this temporal network model was used is that it enables using established seed selection heuristics and diffusion model known for static networks. What is more, the results presented in this work are extending the work [30] by relaxing a single dimension - time. This makes the comparison between applied seeds' activation approaches easier to follow when analysing their performance in static and temporal setting across this and aforementioned work. Formally, the temporal network $TN^K$ can be expressed as a sequence of time-ordered component network snapshots $T_n$ already defined, such that $TN^K = (T_1, \ldots, T_p, \ldots, T_K)$, $K \in \mathbb{N}_+$.

In order to find an in-depth justification for using temporal networks instead of static networks, the reader is advised to look into [44]. However, for this study we also quantified the average stability regarding the nodes and edges for selected number of snapshots. The procedure for computing these numbers was the following. For each network snapshot we did compute the Jaccard index between the sets of nodes and edges in a given snapshot and the one preceding it. This has been done for all snapshots and averaged. This averaged Jaccard index for manufacturing company dataset for sixteen windows was 0.42 for nodes and 0.27 for edges. For thirty two windows it was 0.15 and 0.05, respectively. For Haggle, the stability was lower - for sixteen windows it was 0.28 for nodes and 0.12 for edges, whilst for thirty two windows - 0.07 and 0.04, respectively. This demonstrates that the variability of nodes and edges is high across the snapshots and has significant implications for the spread of influence in terms of how it develops compared to static networks.

### C. SEED SELECTION

Before activating the seeds, the seed set has to be generated. Ideally, seed selection techniques construct this set as the one that has the highest potential for influencing nodes in the network. Due to the hardness of this problem, this implies using heuristics, either basing on the network structure or by exploiting the attributes of nodes [33]. As the purpose of this work is not to evaluate the performance of particular seed selection strategies, but to investigate the capabilities of sequential seeding in a temporal network setting, we did choose a single heuristic that is based on a degree of nodes. As the method introduced in this work is not a seeding strategy but a seed activation method, it can base on different heuristics, including the ones that incorporate more information on the influence spread process, such as the ones presented in [29], [45].

The degree-based heuristic works as follows. For the first snapshot of the network, $T_1$, we compute the degree of all nodes and create an ordered ranking list of nodes, with the highest degree on the top. In case of two nodes having the same degree, the node with the lower id is higher in the ran (e.g. if nodes 7 and 8 have the same degree, node 7 will be before node 8 in the ordered list). For single stage seeding we simply choose top *l* nodes from the list, while for sequential seeding for each seeding step we choose the first inactive node from the list (for details see Section III-E). Please note, that since this is a temporal network, it is not guaranteed that all of the nodes chosen as seeds will appear in further snapshots.

### D. SOCIAL INFLUENCE MODEL

The model chosen for social influence spreading is *independent cascade IC model* [31]. This model assumes that a node has a single chance of activating its neighbour expressed as a probability *p*. If the activation will succeed, this neighbour will become activated and will be attempting to activate its neighbours in the following iterations. As the basic version of this model was proposed for static networks, in the temporal setting we added two modifications to the model:

1) *exhaustion of spreading capabilities* - in every snapshot $T_n$ the iterations follow until no further activations are possible,
2) *single attempt of activation* - if a node failed to activate its neighbour, in the subsequent snapshots it would not be able to try again.

These extensions to the base IC model allow to adequately spread activations over the temporal social network, but at the same time - restrict from reaching all the nodes too early.

As the independent cascade model is not deterministic, for each parameter combination we run 10,000 simulations of the diffusion process. Moreover, in order to make results comparable, we followed the coordinated execution procedure described in [32] - the independence cascade model drawings' results have been the same for all the runs. Thanks to that, the results of the performance of each seeding method are comparable.

### E. SINGLE STAGE SEEDING AND SEQUENTIAL SEEDING

In these experiments two approaches of seeds activation have been evaluated:

- *single stage seeding SGL* - activating all the nodes at the beginning of the diffusion process,
- *sequential seeding SEQ* - activating $j$ nodes in each time window.

Single stage seeding method simply activates all the nodes from the seed set starting from $T_2$ - the first time window for evaluation. If some nodes are not present in $T_2$, they will be activated at the beginning of the window they appear in.

Contrary to that, sequential seeding described in Section II is distributing the activations over subsequent iterations of the process. In the case of temporal network, these will be separate time windows. Following the design of the method, the ordered list of nodes starting from the highest degree that was built as described in Section III-C is used for activating nodes in a manner that if a node was within top $l$ percent of nodes and has not been naturally activated yet, it will become activated. Otherwise, if the natural process of spread of influence led to activation of this node, the next node in list becomes activated. As a result, the same number of activations is being made, but if some original seeds became already activated, new nodes with slightly smaller degree are being activated instead.

The number of seeds to be activated in total, $l$ (see Section III-C is being computed for each combination of number of windows $n$ and number of seeds per window $j$:

$$l = (n - 1) * j \qquad (1)$$

In the above equation, $n$ is decreased by one, because the first window is used for seed selection, while the remaining ones for activations.

More details on the concept of sequential seeding are presented in Section II.

### F. EVALUATION CRITERIA

In a typical influence maximisation problem, the most important factor is what is the total spread of influence for a given method [8], namely how many nodes have been activated at the end of the process. In our work we use the same approach. However, for the temporal network setting, we investigate how many nodes are active at the end of the last snapshot ($T_n$). This means that we sum all the activations in snapshots $T_2 \ldots T_n$, and this sum, usually called total spread of influence, is the base for comparing the effectiveness of the approaches. Apart from that, we also investigate how the process evolved over time to see what how single stage seeding and sequential seeding perform while the process still lasts.

### G. CODE AVAILABILITY

The code for the sequential seeding method in temporal networks and all the experiments conducted in this study

**TABLE 1.** Configuration space of the simulated diffusion processes.

| Parameter | Values | Variants |
|---|---|---|
| Dataset - $d$ | 4 | manufacturing<br>manufacturing-rewired<br>haggle<br>haggle-rewired |
| Propagation probability - $p$ | 9 | 0.05, 0.1, 0.15, 0.2, 0.25, 0.3, 0.4, 0.5, 0.75 |
| Number of windows - $n$ | 5 | 4, 8, 16, 32, 64 |
| Number of seeds per window - $j$ | 5 | 1, 2, 3, 4, 5 |

has been made public and is available as a Code Ocean repository.[1]

## IV. RESULTS

In this section we present the comparison of the performance of single stage seeding SGL and sequential seeding SEQ. We start with the analysis of selected configurations of parameters to show how the spread process develops over time for both approaches. However, in order to generalize the results, Section IV-B summarizes the performance across the parameter space. As described in the previous section, the evaluation criterion is the total number of activations at the end of the last snapshot.

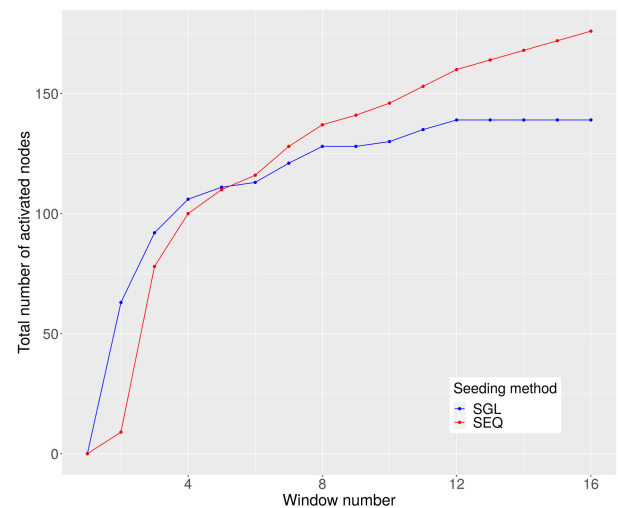

**FIGURE 5.** Seeding strategies over time windows for manufacturing company dataset. Parameters: $p = 0.1, n = 16, j = 3, l = 45$.

### A. SELECTED CASES

In Fig. 5 and 6 the results comparing the performance of evaluated seeding strategies for manufacturing and haggle datasets are presented, respectively. At the beginning of the process, for both datasets single stage seeding gains the advantage due to higher number of initially activated nodes. However, this strategy gets overcome by sequential seeding in the subsequent snapshots due to better spending of the budget for activations. These two figures demonstrate that sequential seeding due to its nature cannot be considered as a method for reaching high spreads in a short time, since this approach is rather for taking the advantage of well-spending the budget based on skipping the nodes that have been naturally activated. This observation has been made for static networks and, as it is demonstrated, also holds for temporal ones.

[1]https://doi.org/10.24433/CO.7254599.v1

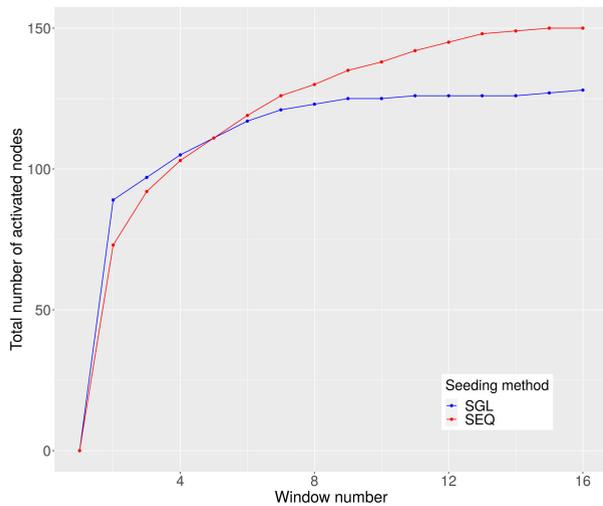

**FIGURE 6.** Seeding strategies over time windows for manufacturing company dataset. Parameters: $p = 0.05, n = 16, j = 4, l = 60$.

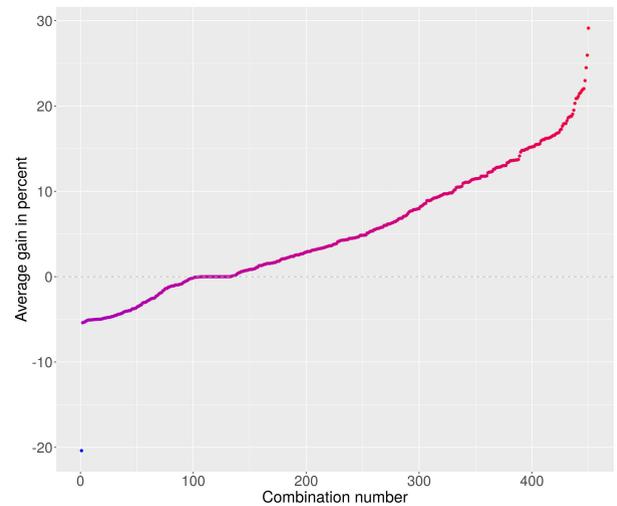

**FIGURE 7.** The percentage gain in terms of number of activated nodes at the end of the spread process for sequential seeding compared to single stage seeding for non-rewired datasets.

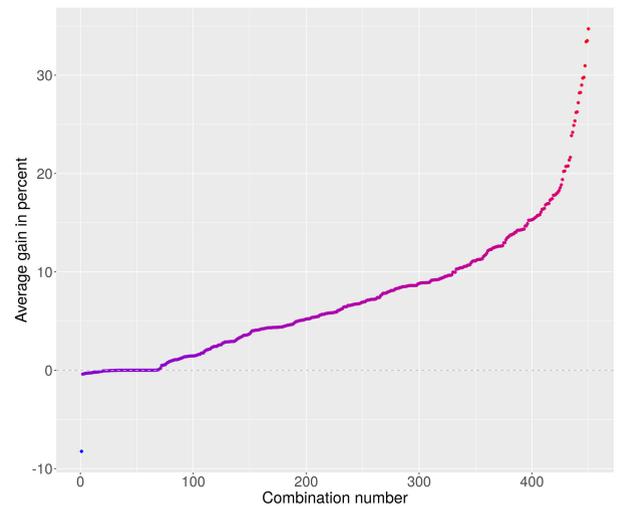

**FIGURE 8.** The percentage gain in terms of number of activated nodes at the end of the spread process for sequential seeding compared to single stage seeding for rewired datasets.

### B. ACTIVATION GAIN

In order to quantify the performance of the sequential seeding against single stage seeding, we computed the gain or loss for all the evaluated combinations of parameters. By these two indicators we mean what is the percentage performance of sequential seeding compared to single stage seeding. This is shown for both types of datasets: the original and rewired ones separately and ordered in increasing order by gain. The visual outcomes of these experiments are presented in Fig. 7 and 8 for non-rewided and rewired network configurations, respectively.

For non-rewired datasets, out of 450 evaluated combinations of parameters, in 318 cases sequential seeding approach outperformed single stage seeding, for 18 cases both methods performed equally and in 114 cases sequential seeding was worse. This means that in nearly 74% out of all combinations sequential seeding performed the same or better compared to single stage seeding. A detailed analysis of the results shows that the median when sequential seeding outperforms single stage seeding equals 7.61% and -2.99% in the opposite case. This leads to the conclusion that the gains not only are more often to be found when exploring the parameter space, but when sequential seeding performs better, the gains are also higher.

In order to find out how the results depend on the causal configuration of edges in temporal datasets, as described in Section III, we also run the experiments using the same datasets, but with edges rewired. In the case of rewired datasets, the sequential seeding approach performs even better, since in 385 combinations of parameters it outperformed single stage seeding, for 40 cases it performed worse and in 25 being equally effective. With respect to the gain, albeit for the case of sequential seeding being better the median is similar and equals 7.20%, for the second case it dropped significantly and equals only -0.07%. When combining the results coming from non-randomized and randomized datasets, in 71% of cases sequential seeding provided better compared to single stage seeding, in 2% being equal in spreading capabilities.

In the next part we focused on analysing the cases in which sequential seeding performed worse than single stage seeding. As it was anticipated, the majority of situations in which sequential seeding performs worse than single stage seeding is related to propagation probability. Namely, if the propagation probability is high (especially 0.25 and above), natural activations tend to overcome the differences in seeding strategies. This results with the situation in which most nodes will be activated at the beginning of the process, resulting in better final outcome. And this is the reason why single stage seeding is superior over sequential seeding for high propagation probabilities. Also, when the probability is less than 0.05, it is too small to let the activations spread over the network and in general in most cases only nodes from the seed set become activated. This is related to the fact that the more time passes from building the seed set, the lower the chance to a node to appear again in a temporal network. Thanks to that, single stage seeding has the opportunity to activate the most of the nodes from the seed set and sequential seeding cannot do so, especially for later snapshots. Another case for lower performance of sequential seeding was observed for high number of snapshots. In Fig. 9 and 10 we investigated at which values of propagation probability sequential seeding is able to outperform single stage seeding. Confirming observations presented above, the highest gain is visible around $p = 10\%$. Similar results have been obtained in the case of static networks [30].

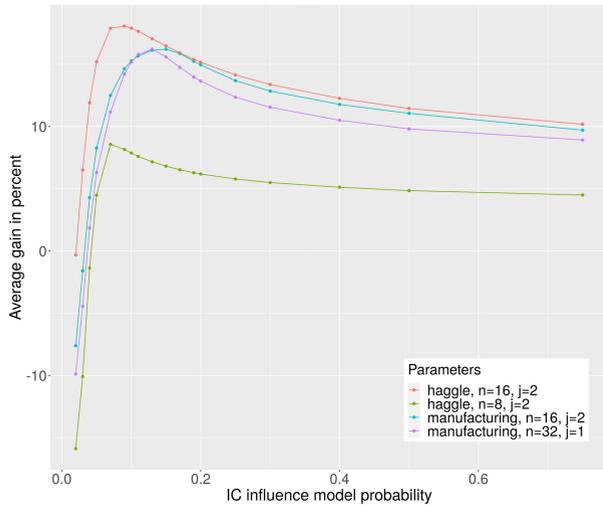

**FIGURE 9.** The percentage gain in terms of number of activated nodes at the end of the spread process for sequential seeding compared to single stage seeding for selected combinations of parameters analysed for higher resolution of probabilities of the influence model.

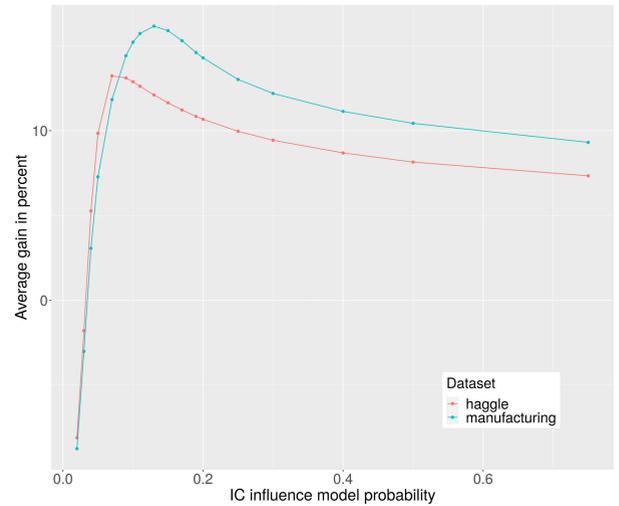

**FIGURE 10.** The percentage gain in terms of number of activated nodes at the end of the spread process for sequential seeding compared to single stage seeding for all combinations of parameters analysed for higher resolution of probabilities of the influence model.

### C. STATISTICAL ANALYSIS

Wilcoxon signed-rank test was used to analyse if there is significant difference between sequential and single stage seeding within temporal network for each propagation probability $p$. It confirms higher performance of the introduced approach (p-value $< 0.05$). The results are presented in Tab. 2. While Wilcoxon signed-rank test confirms statistical significance, the size of the differences between algorithms was represented by Hodges–Lehmann estimator $\Delta$ based on medians of the distributions [46]. Values $\Delta > 0$ demonstrate significantly higher values for coverage of sequential seeding strategy compared to single stage approach. Smallest difference ($\Delta = 4.35$) was observed for lowest propagation probability $p = 0.05$. Then it grows with slight decrease to $\Delta < 7$ for propagation probabilities in range 0.2 - 0.3. Higher propagation probabilities delivered increased performance with $\Delta > 7$. The distance between results for both seeding strategies are consistent with gain presented in Fig. 9 and 10.

**TABLE 2.** Differences between results for single stage and sequential seeding for each propagation probability.

| Propagation probability | Hodges–Lehmann $\Delta$ | p-value |
|---|---|---|
| 0.05 | 4.35 | 0.001544 |
| 0.10 | 7.25 | 0.000024 |
| 0.15 | 7.19 | 0.000028 |
| 0.20 | 6.79 | 0.000043 |
| 0.25 | 6.79 | 0.000040 |
| 0.30 | 6.95 | 0.000052 |
| 0.40 | 7.29 | 0.000041 |
| 0.50 | 7.63 | 0.000033 |
| 0.75 | 7.33 | 0.000071 |

The results presented in Tab. 3 show higher spread in terms of Hodges–Lehmann estimator $\Delta$ for sequential seeding when compared to single stage seeding with the number of windows up to 32. Scenarios with 64 windows delivered worse results for sequential seeding than for single stage seeding. The reason for that is that for high number of windows, only few nodes and edges are to be found in each window. As a result of that, sequential seeding strategy cannot activate nodes, since they are absent in the windows they are about to be activated. This leads to the delays in activations, a problem that is not that much visible for single stage seeding. In the same table, the $p - values$ are presented. They demonstrate the statistical significance for most results apart from 2, 3, 4 seeds used for scenarios with 32 windows and a single seed used for scenario with 64 windows.

**TABLE 3.** The differences between results for single stage and sequential seeding for varying number of windows and number of seeds with positive values for better results for sequential seeding accompanied by the statistical significance of results represented by Hodges–Lehmann estimator (in parentheses).

| Seeds \ Windows | 4 | 8 | 16 | 32 | 64 |
|---|---|---|---|---|---|
| 1 | 1.00 (0.032825) | 3.88 (0.000023) | 9.35 (0.000008) | 10.21 (0.000038) | -2.34 (0.196388) |
| 2 | 2.39 (0.000053) | 8.98 (0.000008) | 18.73 (0.000008) | 4.45 (0.089767) | -3.34 (0.000482) |
| 2 | 4.19 (0.000008) | 14.58 (0.000008) | 21.27 (0.000008) | 4.53 (0.089767) | -2.65 (0.000482) |
| 4 | 6.00 (0.000008) | 20.15 (0.000008) | 17.23 (0.000008) | -1.19 (0.579842) | -1.97 (0.000727) |
| 5 | 7.90 (0.000008) | 23.47 (0.000008) | 11.05 (0.000008) | -5.01 (0.001097) | -1.66 (0.000727) |

## V. CONCLUSION

In this work we introduced and investigated the capabilities of a sequential seeding technique in temporal networks. Compared to the most typical approach, single stage seeding, instead of activating all seed nodes at the same time, sequential seeding distributes activations over time and takes the advantage of observing how the activation process progresses naturally. In our earlier works we demonstrated that sequential seeding is superior to single stage seeding in static networks, but still the question remained whether a temporal setting imposes any change. The results of our experimental study for independent cascades influence model demonstrate that sequential seeding in majority of cases outperforms single stage seeding for real world temporal networks, even when they are rewired. However, this method has limited performance when the propagation probabilities are less than 0.05 or higher than 0.25, since outside this range the dynamics of the influence process reduces the differences between evaluated methods.

It is worth underlining that the sequential seeding approach is not a seed selection heuristic, but seed activation method. As a result, there is still a room for improvement regarding choosing seeding heuristics for influence maximisation in temporal networks. This is being already investigated by researchers [20], but possibly these heuristics combined with sequential seeding can lead to even better results in terms of the spread. This kind of analysis would be a separate study of different purpose and we plan to conduct it in the setting of temporal networks.

The results can be considered as promising ones, especially knowing that there are many challenges in temporal networks that make the problem harder compared to the static scenario. However, the direction toward exploring temporal networks for spread of influence is inevitable, since they are closer to reality, especially when thinking of dynamic processes happening on top of these.

As other future work directions we plan to investigate the capabilities of sequential seeding approach for other social influence models. Another interesting direction would be proposing an optimal strategy for seed activation, as the one evaluated in this work was to equally distribute the seeds over time. Possibly some other approaches based on sequential seeding will lead to even higher gains.

## REFERENCES

[1] A.-L. Barabási and M. Pósfai, *Network Science*. Cambridge, U.K.: Cambridge Univ. Press, 2016.

[2] H. Hexmoor, *Computational Network Science: An Algorithmic Approach*. San Mateo, CA, USA: Morgan Kaufmann, 2014.

[3] D. Lazer, A. Pentland, L. Adamic, S. Aral, A.-L. Barabási, D. Brewer, N. Christakis, N. Contractor, J. Fowler, M. Gutmann, T. Jebara, G. King, M. Macy, D. Roy, and M. Van Alstyne, "Computational social science," *Science*, vol. 323, no. 5915, pp. 721–723, 2009.

[4] K. Kandhway and J. Kuri, "How to run a campaign: Optimal control of SIS and SIR information epidemics," *Appl. Math. Comput.*, vol. 231, pp. 79–92, Mar. 2014.

[5] R. T. Zaman, R. Herbrich, J. V. Gael, and D. Stern, "Predicting information spreading in Twitter," in *Proc. Comput. Social Sci. Wisdom Crowds Workshop (Colocated NIPS)*, Dec. 2010, pp. 1–4.

[6] P. Brodka, K. Musial, and J. Jankowski, "Interacting spreading processes in multilayer networks: A systematic review," *IEEE Access*, vol. 8, pp. 10316–10341, 2020.

[7] X. Chen, R. Wang, M. Tang, S. Cai, H. E. Stanley, and L. A. Braunstein, "Suppressing epidemic spreading in multiplex networks with social-support," *New J. Phys.*, vol. 20, no. 1, Jan. 2018, Art. no. 013007.

[8] D. Kempe, J. Kleinberg, and É. Tardos, "Maximizing the spread of influence through a social network," in *Proc. 9th ACM SIGKDD Int. Conf. Knowl. Discovery Data Mining (KDD)*, 2003, pp. 137–146.

[9] A. Goyal, W. Lu, and L. V. S. Lakshmanan, "CELF++: Optimizing the greedy algorithm for influence maximization in social networks," in *Proc. 20th Int. Conf. Companion World Wide Web (WWW)*, 2011, pp. 47–48.

[10] X. Wang, X. Zhang, C. Zhao, and D. Yi, "Maximizing the spread of influence via generalized degree discount," *PLoS ONE*, vol. 11, no. 10, Oct. 2016, Art. no. e0164393.

[11] Y. Zhao, S. Li, and F. Jin, "Identification of influential nodes in social networks with community structure based on label propagation," *Neurocomputing*, vol. 210, pp. 34–44, Oct. 2016.

[12] C. Lee, F. Reid, A. McDaid, and N. Hurley, "Seeding for pervasively overlapping communities," *Phys. Rev. E, Stat. Phys. Plasmas Fluids Relat. Interdiscip. Top.*, vol. 83, no. 6, Jun. 2011, Art. no. 066107.

[13] M. Kitsak, L. K. Gallos, S. Havlin, F. Liljeros, L. Muchnik, H. E. Stanley, and H. A. Makse, "Identification of influential spreaders in complex networks," *Nature Phys.*, vol. 6, no. 11, p. 888, 2010.

[14] O. Hinz, B. Skiera, C. Barrot, and J. U. Becker, "Seeding strategies for viral marketing: An empirical comparison," *J. Marketing*, vol. 75, no. 6, pp. 55–71, Nov. 2011.

[15] F. Erlandsson, P. Bródka, and A. Borg, "Seed selection for information cascade in multilayer networks," in *Proc. Int. Conf. Complex Netw. Appl.* Cham, Switzerland: Springer, 2017, pp. 426–436.

[16] J.-X. Zhang, D.-B. Chen, Q. Dong, and Z.-D. Zhao, "Identifying a set of influential spreaders in complex networks," *Sci. Rep.*, vol. 6, no. 1, Jun. 2016, Art. no. 27823.

[17] J. C. S. Freitas, R. Barbastefano, and D. Carvalho, "Identifying influential patents in citation networks using enhanced VoteRank centrality," 2018, *arXiv:1811.01638*. [Online]. Available: http://arxiv.org/abs/1811.01638

[18] M. M. Tulu, R. Hou, and T. Younas, "Vital nodes extracting method based on user's behavior in 5G mobile social networks," *J. Netw. Comput. Appl.*, vol. 133, pp. 39–50, May 2019.

[19] C. C. Aggarwal, S. Lin, and P. S. Yu, "On influential node discovery in dynamic social networks," in *Proc. SIAM Int. Conf. Data Mining*, Apr. 2012, pp. 636–647.

[20] R. Michalski, T. Kajdanowicz, P. Bródka, and P. Kazienko, "Seed selection for spread of influence in social networks: Temporal vs. static approach," *New Gener. Comput.*, vol. 32, nos. 3–4, pp. 213–235, Aug. 2014.

[21] C.-T. Li, H.-P. Hsieh, S.-D. Lin, and M.-K. Shan, "Finding influential seed successors in social networks," in *Proc. 21st Int. Conf. Companion World Wide Web*, 2012, pp. 557–558.
[22] J. Jankowski, R. Michalski, and P. Kazienko, "Compensatory seeding in networks with varying avaliability of nodes," in *Proc. IEEE/ACM Int. Conf. Adv. Social Netw. Anal. Mining (ASONAM)*, 2013, pp. 1242–1249.
[23] A. Wald, *Sequential Analysis*. Chelmsford, MA, USA: Courier Corporation, 2004.
[24] R. M. Cyert, M. H. DeGroot, and C. A. Holt, "Sequential investment decisions with Bayesian learning," *Manage. Sci.*, vol. 24, no. 7, pp. 712–718, Mar. 1978.
[25] W. H. Price, "Sequential immunization as a vaccination procedure against dengue viruses," *Amer. J. Epidemiol.*, vol. 88, no. 3, pp. 392–397, 1968.
[26] A. Elberse and J. Eliashberg, "Demand and supply dynamics for sequentially released products in international markets: The case of motion pictures," *Marketing Sci.*, vol. 22, no. 3, pp. 329–354, Aug. 2003.
[27] A. Sela, I. Ben-Gal, A. S. Pentland, and E. Shmueli, "Improving information spread through a scheduled seeding approach," in *Proc. IEEE/ACM Int. Conf. Adv. Social Netw. Anal. Mining (ASONAM)*, Aug. 2015, pp. 629–632.
[28] A. Sela, D. Goldenberg, I. Ben-Gal, and E. Shmueli, "Active viral marketing: Incorporating continuous active seeding efforts into the diffusion model," *Expert Syst. Appl.*, vol. 107, pp. 45–60, Oct. 2018.
[29] Y. Wang, J. Zhu, and Q. Ming, "Incremental influence maximization for dynamic social networks," in *Proc. Int. Conf. Pioneering Comput. Sci., Eng. Educators*. Singapore: Springer, 2017, pp. 13–27.
[30] J. Jankowski, P. Bródka, P. Kazienko, B. K. Szymanski, R. Michalski, and T. Kajdanowicz, "Balancing speed and coverage by sequential seeding in complex networks," *Sci. Rep.*, vol. 7, no. 1, Dec. 2017, Art. no. 891.
[31] J. Goldenberg, B. Libai, and, E. Muller, "Talk of the network: A complex systems look at the underlying process of word-of-mouth," *Marketing Lett.*, vol. 12, no. 3, pp. 211–223, 2001.
[32] J. Jankowski, B. K. Szymanski, P. Kazienko, R. Michalski, and P. Bródka, "Probing limits of information spread with sequential seeding," *Sci. Rep.*, vol. 8, no. 1, Dec. 2018, Art. no. 13996.
[33] R. Michalski and P. Kazienko, "Maximizing social influence in real-world networks—The state of the art and current challenges," in *Propagation Phenomena in Real World Networks*. Cham, Switzerland: Springer, 2015, pp. 329–359.
[34] M. Weskida and R. Michalski, "Finding influentials in social networks using evolutionary algorithm," *J. Comput. Sci.*, vol. 31, pp. 77–85, Feb. 2019.
[35] J. Jankowski, P. Bródka, R. Michalski, and P. Kazienko, "Seeds buffering for information spreading processes," in *Proc. Int. Conf. Social Informat.* Cham, Switzerland: Springer, 2017, pp. 628–641.
[36] J. Jankowski, "Dynamic rankings for seed selection in complex networks: Balancing costs and coverage," *Entropy*, vol. 19, no. 4, p. 170, Apr. 2017.
[37] Q. Liu and T. Hong, "Sequential seeding for spreading in complex networks: Influence of the network topology," *Phys. A, Stat. Mech. Appl.*, vol. 508, pp. 10–17, Oct. 2018.
[38] J. Jankowski, M. Waniek, A. Alshamsi, P. Bródka, and R. Michalski, "Strategic distribution of seeds to support diffusion in complex networks," *PLoS ONE*, vol. 13, no. 10, Oct. 2018, Art. no. e0205130.
[39] C. Ni, J. Yang, and D. Kong, "Sequential seeding strategy for social influence diffusion with improved entropy-based centrality," *Phys. A, Stat. Mech. Appl.*, vol. 545, May 2020, Art. no. 123659.
[40] M. Nurek and R. Michalski, "Combining machine learning and social network analysis to reveal the organizational structures," *Appl. Sci.*, vol. 10, no. 5, p. 1699, Mar. 2020.
[41] A. Chaintreau, P. Hui, J. Crowcroft, C. Diot, R. Gass, and J. Scott, "Impact of human mobility on opportunistic forwarding algorithms," *IEEE Trans. Mobile Comput.*, vol. 6, no. 6, pp. 606–620, Jun. 2007.
[42] L. Gauvin, M. Génois, M. Karsai, M. Kivelä, T. Takaguchi, E. Valdano, and C. L. Vestergaard, "Randomized reference models for temporal networks," 2018, *arXiv:1806.04032*. [Online]. Available: http://arxiv.org/abs/1806.04032
[43] D. Kempe, J. Kleinberg, and É. Tardos, "Maximizing the spread of influence through a social network," *Theory Comput.*, vol. 11, no. 1, pp. 105–147, 2015.
[44] P. Holme and J. Saramäki, "Temporal networks," *Phys. Rep.*, vol. 519, no. 3, pp. 97–125, 2012.
[45] X. Liu, X. Liao, S. Li, S. Zheng, B. Lin, J. Zhang, L. Shao, C. Huang, and L. Xiao, "On the shoulders of giants: Incremental influence maximization in evolving social networks," *Complexity*, vol. 2017, pp. 1–14, Sep. 2017.
[46] E. L. Lehmann and H. J. D'Abrera, *Nonparametrics: Statistical Methods Based on Ranks*. Queanbeyan, NSW, Australia: Holden-Day, 1975.

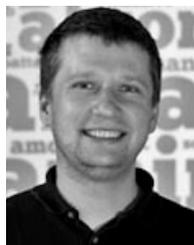

**RADOSŁAW MICHALSKI** (Member, IEEE) is currently an Assistant Professor with the Department of Computational Intelligence, Wrocław University of Science and Technology, Wrocław, Poland. His research interests include social influence, diffusion processes in complex networks, and machine learning. He has coauthored over 50 publications in these areas, ten of which were published in influential journals.

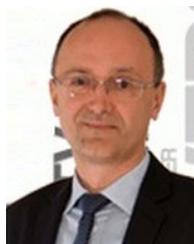

**JAROSŁAW JANKOWSKI** currently works as an Associate Professor with the Faculty of Computer Science, West Pomeranian University of Technology, Poland. His research interests include the interdisciplinary field of human centered computing with focus on user behavior analysis, agent-based modeling, and diffusion of information in social networks. He is also working on the integration of computer science methods, with social aspects of computer systems, with results published in journals like *Scientific Reports*, *Computers in Human Behavior*, and the *Journal of Management Information Systems*, and presentations related to viral marketing, social networks, and human–computer interaction on various international conferences.

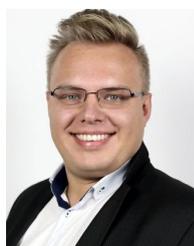

**PIOTR BRÓDKA** received the M.Sc. degree in computer science from the Wrocław University of Technology, Poland, in 2008, and the Ph.D. degree in 2012. He was a Visiting Scholar with Stanford University, in 2013, and a Visiting Professor with the University of Technology Sydney, in 2018 and 2019. He is currently an Assistant Professor of computer science with the Department of Computational Intelligence, Wrocław University of Science and Technology. He has authored over 80 scholarly and research articles on a variety of areas related to complex networks, focusing on the extraction and dynamics of communities within social networks, spreading processes in complex networks, and the analysis of multilayer networks. In 2015, he has received three years Scholarship for the Best Young Scientists awarded by the Polish Ministry of Science and Higher Education. He has attended many conferences on complex networks and computational network science.